\documentclass[%
 aip,
 amsmath,amssymb,
 reprint,%
]{revtex4-1}

\usepackage{chemformula} 
\usepackage[T1]{fontenc} 

\usepackage{amssymb}
\usepackage{multirow}
\usepackage{bbold}

\newcommand{\kSD}{k_{SD}}
\newcommand{\kRbf}{k_{RBF}}
\newcommand{\kRQ}{k_{RQ}}
\newcommand{\kMat}{k_{^{MAT}_{5/2}}}
\newcommand{\kMatt}{k_{^{MAT}_{3/2}}}
\newcommand{\dSD}{|\delta|}

\draft 

\begin{document}


\title{Gaussian Processes with Spectral Delta kernel for higher accurate Potential Energy surfaces for large molecules} 



\author{Rodrigo A. Vargas--Hern\'andez}
\email{r.vargashernandez@utoronto.ca}
\affiliation{Chemical Physics Theory Group, Department of Chemistry, University of Toronto,Toronto,Ontario M5S 3H6, Canada.}
\affiliation{Vector Institute for Artificial Intelligence, Toronto, Canada.}

\author{Jake R. Gardner}
\affiliation{Dept. of Computer and Information Science, University of Pennsylvania, Philadelphia, PA, USA. }


\date{\today}

\begin{abstract}
The interpolation of high-dimensional potential energy surfaces (PESs) is commonly done with physically-inspired deep-neural network models.
In this work, we illustrate that Gaussian Processes (GPs) are also capable of interpolating high-dimensional complex physical systems.
The accuracy of GPs depends on the robustness of the kernel function, and a boost in the accuracy is achieved by linearly combining kernel functions. 
In this work, we proposed an alternative route by parametrizing the kernel function through Bochners’ theorem.
We interpolated the PES of various chemical systems achieving a global accuracy of $<$ 0.06 kcal/mol for Benzene, Malonaldehyde, Ethanol, and protonated Imidazole dimer using only 15 000 training points. Additionally, for Aspirin, we achieved a global error of 0.063 kcal/mol with 20 000 points.
Given these results, we believe this kernel function is system-agnostic and could allow GPs to tackle a wider variety of high-dimensional physical systems.
\end{abstract}


\maketitle 

\section{Introduction}
Interpolation of potential energy surfaces (PESs) is one of the most common applications of supervised machine learning (ML) methods.
Commonly, this is done with deep neural networks (DNNs) \cite{GP_vs_NN,NN_Tucker,NN_small_molecules,NN_highdim,many-body_NN,NN_environment-dependent_atom,NN_ANI-1,DTNN,SchNet,Yafu_diab_NN,Cormorant,PhysNet,PCA_NN_PES}, parametric models, or kernel models, such as Gaussian Processes (GPs) or kernel ridge regression (KRR)  \cite{GP_PRL,GP_vs_NN,RAVH_NJP,GP_NaKNak_JCP,Jun_JCTC,Hiroki_JCP,Qingyong_GP,Multi-fidelity_GP,Gauss_PES,GP_Cui_JPhysB,GP_reactive_PES,GP_reactive_PES_2,GP_PES-Learn}. 
While both methodologies have proven to be flexible enough, GPs require less "tunning" compared to NNs, where the search for an optimal architecture could be computationally demanding \cite{NN_NAS_2019,NN_NAS_2020}. 
One of the advantages of GPs is their ability to quantify the uncertainty in their prediction, which is commonly used in applications where noisy data or the optimization of $f$ black-box functions. The latter application has proven to be a successful tool in physical-chemistry \cite{RAVH_NJP,RAVH_BODFT,Deng_BO_JCP,Phoenics_BO,BO_DFT_water,BO_Materials,MOBOpt_vargas}.\\

GPs and KRR mainly depend on two factors: training data, and the kernel function. The latter quantifies the similarity between a pair of points. 
If the kernel function can generalize a similarity metric, GPs are efficient regression algorithms that require less training data than NNs \cite{GP_vs_NN}, and are also capable of extrapolating functions beyond the training data regime \cite{RAVH_PRL,RAVH_extrapolation,Jun_JCTC,duvenaud_bic,Krems_PRE}. To achieve more robust kernel functions, once can simply combine different simple kernel functions \cite{duvenaud_kerneladdition,gpbook}. However, to automate the search for an optimal kernel combination, one can use Bayesian model selection to search over possible priors, where a prior is determined by the kernel function form. Given the combinatorial nature of the space of possible kernel functions, sampling this distribution is hard and one must rely on heuristic search approaches. One such approach is the so called Automatic Bayesian Covariance Discovery (ABCD) \cite{duvenaud_bic,duvenaud_kerneladdition}, described in greater detail below. Recently, people have proposed the use of Bayesian optimization for this as well \cite{malkomes2016bayesian}. \\

Another technique that has proven to construct accurate PES is based on KRR using the atomic gradient information, also known as gradient-domain machine learning (GDML) \cite{GDML_1,GDML_2}.
Here, the kernel complexity is usually not modified. However, a combination of kernels could be easily implemented given the analytic derivatives of them \cite{GDML_Krems}; however, the training procedure is based on all possible combinations of kernels and the selection of them is through a cross-validation scheme.
The computational complexity of GDML also scales cubically with the number of training data.
In some cases, for large covariance matrices the Nystr\"om approximation is applied \cite{GP_BigData,GP_BigData_Rev}.\\ 

Although kernel structure search has been shown to accurately interpolate of PESs, they are not often applied to study high-dimensional physical systems due to a lack of accuracy and significant computational cost of ${\cal O}(N^3)$, where $N$ is the number of training points. \cite{GP_NaKNak_JCP,Jun_JCTC,Hiroki_JCP,Qingyong_GP}. Typically, approximations must be used to achieve the required scalability, which can degrade the predictive performance of GP models even further. For systems that require a higher number of training data points, multiple GPs must be trained in order to find the optimal kernel combination.

In this work, we present a different approach to construct kernel functions based on Bochners’ theorem \cite{gpbook} that are both more robust and computationally efficient.\\ In the following sections, we briefly introduce GPs and review how complex kernels can be constructed by using Bochners’ theorem, a strategy known as \emph{spectral kernel learning}.
We interpolate the PES of four different chemical systems, e.g., protonated Imidazole dimer \cite{Hiroki_JCP}, Benzene , Malonaldehyde, and Ethanol \cite{GDML_1,GDML_2}.
We also illustrate that by using GPytorch \cite{gpytorch} and KEOPS \cite{keops}, modern deep-learning libraries, combined with GPUs we can efficiently train full GPs with 20 000 training points without relying on the Nystr\"om approximation. 

\section{Gaussian Processes}
Gaussian Processes are one of the most used non-parametric probabilistic ML models in physical sciences. 
A GP is specified by its mean  ($\mu$) and covariance ($\sigma$) functions, $f(\mathbf{x}) \sim {\cal GP}(\mu,\Sigma)$.
One of the key components in a GP model is the kernel function, $k(\mathbf{x}_i,\mathbf{x}_j)$, used to parametrized the covariance matrix. 
For prediction, GPs have an exact posterior distribution whose mean and standard deviation are,
\begin{eqnarray}
\mu(\mathbf{x^*}) &=& \mathbf{k}(\mathbf{x^*},\mathbf{X})^\top \left [ K(\mathbf{X},\mathbf{X}) + \sigma_n 1 \right]^{-1} \mathbf{y} \label{eqn:gp_mu}\\
\sigma(\mathbf{x^*}) &=& k(\mathbf{x^*},\mathbf{x^*}) \left [ K(\mathbf{X},\mathbf{X}) + \sigma_n 1 \right]^{-1} \mathbf{k}(\mathbf{x^*},\mathbf{X}),\label{eqn:gp_cov}
\end{eqnarray}
where $\mathbf{X}$, for PESs, contains the geometry configurations and $\mathbf{y}$ is the energy for each configuration. $K(\cdot,\cdot)$ is known as the design or covariance matrix, whose elements are given by the kernel function, $K_{i,j} = k(\mathbf{x}_i,\mathbf{x}_j)$. $\mathbf{x}^*$ is the point where prediction is aimed, and $\mathbf{k}(\mathbf{x^*},\mathbf{X})$ is the similarity vector between $\mathbf{x^*}$ and all the training data.\\

Given the nature of GPs, it is possible to find the marginal distribution in closed form, $p(\mathbf{y}|\mathbf{X})$. It is common to use the logarithm of the marginal likelihood distribution, LMLL, to optimize the free-parameters of the kernel function ($\mathbf{\theta}$), 
\begin{eqnarray}
\log p(\mathbf{y}|\mathbf{X},\mathbf{\theta}) &=& -\frac{1}{2}\mathbf{y}^\top\left [ K(\mathbf{X},\mathbf{X}) + \sigma_n \mathbb{1} \right]^{-1}\mathbf{y} \nonumber \\ &&- \frac{1}{2}\log \Big | K(\mathbf{X},\mathbf{X}) + \sigma_n \mathbb{1} \Big| -\frac{N}{2}\log 2\pi.
\label{eqn:lmll}
\end{eqnarray}
Some of the most common kernel functions are, 
\begin{eqnarray}
k_{RBF}(\mathbf{x}_i,\mathbf{x}_j) &=& \exp \Big [ -\frac{1}{2}r^{2}(\mathbf{x}_i,\mathbf{x}_j) \Big] \label{eqn:k_rbf}\\
k_{RQ}(\mathbf{x}_i,\mathbf{x}_j) &=& \left ( 1 + \frac{|\mathbf{x}_i - \mathbf{x}_j|^2}{2\alpha\ell^2}\right )^{-\alpha} \label{eqn:k_rq}\\
k_{^{MAT}_{3/2}}(\mathbf{x}_i,\mathbf{x}_j) &=& \left (1 + \sqrt{3}r(\mathbf{x}_i,\mathbf{x}_j) \right ) \exp \Big [ -\sqrt{3}r(\mathbf{x}_i,\mathbf{x}_j) \Big] \nonumber \\ \label{eqn:k_mat_32}\\
k_{^{MAT}_{5/2}}(\mathbf{x}_i,\mathbf{x}_j) &=& \left (1 + \sqrt{5}r(\mathbf{x}_i,\mathbf{x}_j) + \frac{5}{3} r^{2}(\mathbf{x}_i,\mathbf{x}_j) \right ) \nonumber \\ &&\times \exp \Big [ -\sqrt{5}r(\mathbf{x}_i,\mathbf{x}_j) \Big] \label{eqn:k_mat}
\end{eqnarray}
where $\mathbf{r}^2(\mathbf{x}_i,\mathbf{x}_j) = (\mathbf{x}_i - \mathbf{x}_j)^\top M (\mathbf{x}_i - \mathbf{x}_j)$, where $M$ is a diagonal matrix parametrized with a $\ell_d$ length scale for each dimension of $\mathbf{x}$.
For more details about GPs we refer the reader to Ref. \citenum{gpbook}.

In previous works, it has been shown the accuracy of GPs increase by simply combining kernels to account for more complicated functions. One of the most common approaches was proposed in Refs. \citenum{duvenaud_kerneladdition,duvenaud_bic,RAVH_PRL}, where the combination of kernels was guided by selecting the kernel that has the maximum LMLL. To avoid selecting a kernel (${\cal M}_i$) with a large number of parameters, a term that penalizes the number of free parameters ($|{\cal M}_i|$) in the kernel was included, 
\begin{eqnarray}
\text{BIC}({\cal M}_i) = \log p(\mathbf{y}|\mathbf{X},\mathbf{\theta}, {\cal M}_i) + \frac{1}{2} |{\cal M}_i|\log N. 
\label{eqn:BIC}
\end{eqnarray}
This methodology is commonly named as the \emph{Bayesian information criterion} (BIC). 
These kernel structure discovery methods have been demonstrated to extrapolate physical observables accurately enough to detect phase transitions \cite{RAVH_PRL,RAVH_extrapolation}.
Additionally, GPs with complex kernels have shown the possibility to predict accurate energies for PES trained only with low energy points \cite{Jun_JCTC,Hiroki_JCP}.
The computational complexity of Gaussian processes compounded with the fact that kernel search requires training many such models has resulted in an inability to use full available datasets \cite{Jun_JCTC,Hiroki_JCP,Qingyong_GP}.
Also, by considering highly complex kernel combinations, the optimization of the kernel parameters becomes harder, and given the greedy search strategy in practice one could end up with non-optimal kernels. 
Finally, raising the complexity of the kernel combination in practice hits a plateau in the learning capacity meaning, adding more kernels do not increase the accuracy of the model \cite{Jun_JCTC}.\\

Another possibility to construct kernel functions automatically is through Bochners' theorem \cite{gpbook},  
\begin{eqnarray}
k(\tau) = \int_{\mathbb{R}^{D}} e^{2\pi i s^\top\tau} S(\mathbf{s}) \;\mathrm{d} \mathbf{s},
\label{eqn:Bochners_theorem}
\end{eqnarray}
 where $\mathbf{\tau} = \mathbf{x}_i - \mathbf{x}_j$ and $S(s)$ is the spectral density of the $k(\cdot,\cdot)$. $S(\cdot)$ and $k(\cdot)$ are Fourier duals.
 In order for $k(\cdot)$ to be a valid kernel, $S(\cdot)$ must be integrable. For example, the spectral density of the SE kernel is also a Gaussian function.
This theorem ensures that the kernel $k(\tau)$ parametrizes a positive-definite covariance matrix.
The Matern family of covariance functions can also be derived using the Bochners' theorem \cite{gpbook}.
 
 In Ref. \citenum{GP_SD}, L\'{a}zaro-Gredilla \emph{et al.} proposed a novel way to construct $S(\mathbf{s})$ by assuming it is proportional to a probability measure, $S(\mathbf{s})\propto p_{S}(\mathbf{S})$.
 By doing so, the integral over the frequency domain can be computed by Monte Carlo,
 \begin{eqnarray}
 k(\mathbf{x}_i - \mathbf{x}_j) &=& \int_{\mathbb{R}^{D}} \; \mathrm{d} \mathbf{s} \;\; e^{2\pi i \mathbf{s}^\top\left( \mathbf{x}_i - \mathbf{x}_j\right )} S(\mathbf{s}) \nonumber \\ &=& \sigma_0^2 \int_{\mathbb{R}^{D}} \; \mathrm{d} \mathbf{s} \;\; e^{2\pi i \mathbf{s}^\top \mathbf{x}_i}  e^{-2\pi i s^\top \mathbf{x}_j} p_{S}(\mathbf{s}) \nonumber \\
 &=& \sigma_0^2 \mathbb{E}_{p_{S}}\left [ e^{2\pi i \mathbf{s}^\top \mathbf{x}_i}  e^{-2\pi i \mathbf{s}^\top \mathbf{x}_j} \right],
 \label{eqn:Bochners_theorem_2}
 \end{eqnarray}
 where $\sigma_0^2$ is a normalization constant, and $\mathbb{E}_{p_{S}}$ is the expectation value with respect to $p_{S}$. The samples from $p_{S}$ used to compute $k(\cdot,\cdot)$ are known as \emph{spectral points}.
 
\begin{figure}[h!]
\centering
\includegraphics[width=0.45\textwidth]{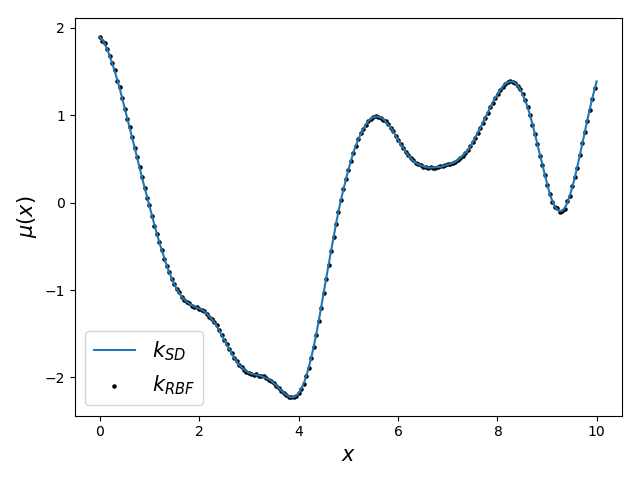}
\includegraphics[width=0.45\textwidth]{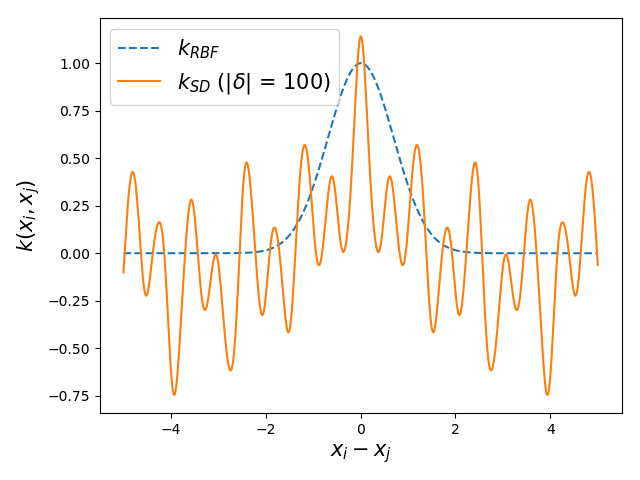}
\caption{(left panel) We sample a GP with the $\kRbf$, black symbols, and train a second GP where the covariance function is $\kSD$ with $\dSD = 100$. The mean of this GP, Eq. \ref{eqn:gp_mu}, is the blue solid curve. (right panel) Dashed blue curve is the target covariance function, $\kRbf$, where data was sampled, and the solid curve is the approximated one with the $\kSD$ kernel. We optimize the parameters of $\kSD$ by maximizing the LMLL, Eq. \ref{eqn:lmll}. 
}
\label{fig:gp_sd}
\end{figure}

 It is possible to cancel the imaginary part of $k(\cdot,\cdot)$ in Eq. \ref{eqn:Bochners_theorem_2} by sampling a pair of $\{\mathbf{s}_r,-\mathbf{s}_r\}$. This Monte Carlo procedure is valid given that $S(\mathbf{s})$ is symmetric around zero. 
 By taking this into account, $k(\cdot,\cdot)$ has the following closed form,
 \begin{eqnarray}
  k(\mathbf{x}_i - \mathbf{x}_j) &\simeq& \frac{\sigma_0^2}{2|\delta|}\sum_{r=1}^{|\delta|} \left [e^{2\pi i \mathbf{s}_r^\top \mathbf{x}_i}  e^{-2\pi i s_r^\top \mathbf{x}_j} + e^{-2\pi i \mathbf{s}_r^\top \mathbf{x}_i}  e^{2\pi i s_r^\top \mathbf{x}_j} \right] \nonumber \\
  &=& \frac{\sigma_0^2}{|\delta|}\sum_{r=1}^{|\delta|} \cos(2\pi\mathbf{s}_r^\top(\mathbf{x}_i-\mathbf{x}_j)),
  \label{eqn:k_SD}
 \end{eqnarray}
 where $|\delta|$ is the total number of spectral points or samples used to approximate Eq. \ref{eqn:Bochners_theorem_2}. $\mathbf{s}_r$ are the frequencies that will be learned by maximizing Eq. \ref{eqn:lmll}. Because this kernel function corresponds to an explicit finite basis expansion,
 inference can be done in ${\cal O}(N)$ time and space \cite{gpytorch}.
As it is stated in Ref. \citenum{GP_SD}, this approximation is similar to a set of Dirac deltas with amplitude $\sigma_0^2$ which are distributed accordingly to $p(\mathbf{s})$. For this work, we denote this kernel as the \emph{spectral delta} kernel, $k_{SD}(\cdot,\cdot)$. Figure \ref{fig:gp_sd} depicts a simple example on how to approximate $\kRbf$ with 100 spectral points.
It should be noted that $k_{SD}(\cdot,\cdot)$ is not the only possible kernel that can be derived from the Bochners' theorem. 
In Ref. \citenum{GP_SMK}, it was shown that by assuming $S(\mathbf{s})$ as a linear combination of two Gaussians centered at $\mathbf{s}$ and $-\mathbf{s}$, the integral in the Fourier space has a close form where $k(\mathbf{\tau}) = \sum_{q=1}^{Q}\omega_q e^{-2\pi\mathbf{\tau}^\top M_q \tau} \cos(2\pi\mathbf{\tau}^\top\mu_q)$.  $\omega_q$, $M_q$ and $\mu_q$ are the free parameters of this kernel.

\section{Results and discussion}
\begin{figure}[h!]
\centering
\includegraphics[width=0.5\textwidth]{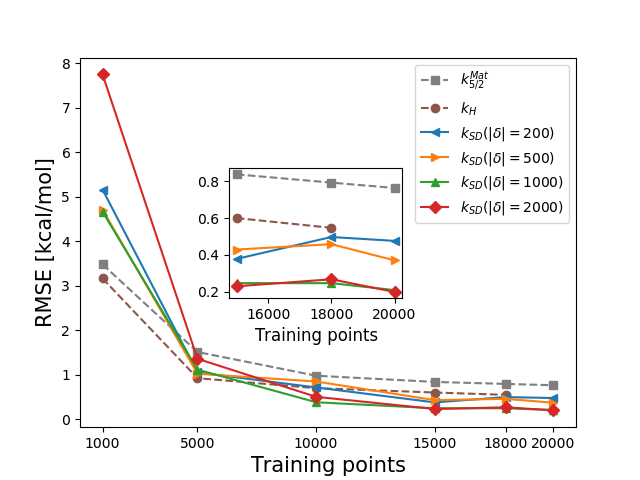}
\caption{RMSE of the protonated Imidazole dimer interpolated with a GP with different kernels. We considered the $\kMat$ (Eq. \ref{eqn:k_mat}), $k_{H}$ (Eq. \ref{eqn:k_Hiroki_simple}), and $\kSD$ (Eq. \ref{eqn:k_SD}) with a different number of $\dSD$. 
The RMSE for each model was computed with 10 000 energy points that were not including in the training set. 
}
\label{fig:Imidazole_simple}
\end{figure}

Here, we first compare the accuracy of the PES for the protonated imidazole dimer  \cite{Hiroki_JCP} interpolated with GPs with various kernel functions,  i.g., Matern (Eq. \ref{eqn:k_mat}), Spectral Delta (Eq. \ref{eqn:k_SD}), and kernel combination through the BIC method.
For the latter, we used the kernel that was optimized through the BIC method,  with 5 000 points, in Ref. \citenum{Hiroki_JCP}, 
\begin{eqnarray}
 k_{H}(\mathbf{x}_i,\mathbf{x}_j) = a_0 \kMat + a_1\kMatt + a_2\kRbf + a_3\kRbf, \nonumber \\
 \label{eqn:k_Hiroki_simple}
\end{eqnarray}
where all $a_i$s and the internal parameters of all the kernels were optimized by maximizing Eq. \ref{eqn:lmll}. 
The PES for the protonated imidazole dimer is a 51D surface with a range of energy points spanning from [0, 35 000] cm$^{-1}$.
For the $\kSD$, we considered a various number of spectral points, $|\delta|$, to study the impact on the accuracy of the model's prediction. 
 We quantified the accuracy of each model by computing the root-mean-square error (RMSE),
 \begin{eqnarray}
\text{RMSE} = \sqrt{\frac{1}{n} \sum_i^{n} \left ( y_i - \hat{y}_i \right )^2}, \nonumber \\
\label{eqn:RMSE}
\end{eqnarray}
where $y_i$s are the predicted values with each GP, and $\hat{y}_i$ are the exact values computed at the MP2/6-31++G($d,p$) level of theory\cite{Hiroki_JCP}.
The test data set consists of 10 000 points spread throughout the same energy-range of the training points. 
All kernel parameters for each different GP were optimized by maximizing LMLL in the GPytorch\cite{gpytorch} suite using the Adam optimizer\cite{Adam}.
To speed up the computation of the GPs with the $\kMat$ and $k_{H}$ kernels we used the KEOPS library\cite{keops}.  
All calculations were carried in a single GPU, V100 SXM2 32GB. 
For a GP with $k_{H}$, 18 000 training points was the maximum number of points that we could used before running out of memory. 
The RMSEs for all different models, as a function of training points, are depicted in Fig. \ref{fig:Imidazole_simple}. 

We found that in the low limit of training data, a GP with a simple kernel can predict a more accurate PES than a GP with the $\kSD$ kernel regardless of the number of spectral points.
However, as the number of training points increases so does the accuracy of a GP with $\kSD$.
This correlates with the idea that more points contain more information therefore more robust kernels can be designed. 
There is a significant difference between a $\kSD$ with 500 and 1 000 spectral points, where the RMSE is almost half of the most simple $\kSD$.
The most accurate PES was achieved with a GP trained with 20 000 points using the $\kSD$ kernel with 5 000 spectral points, RMSE = 0.176 kcal/mol; twice more accurate than a GP with the $k_{H}$ kernel and 18 000 points (RMSE = 0.547 kcal/mol).

Following the same procedure as in Ref. \citenum{Hiroki_JCP}, we split the total PES for Imidazole dimmer into three different PESs, one for each fragment of the system. 
By fragmenting the entire PES, a more accurate model is achieved. 
We denote each individual fragment as $\boldsymbol{\xi}_{i} \sim {\cal GP}(\mu,k_{ij})$. 
In Table \ref{tbl:Imidazole}, we report the error for each fragment and the global PES. We compare the results produce with $\kSD$ with a GP with the $\kMat$ kernel and the ones optimized with the BIC method \cite{Hiroki_JCP}.
For fragments 1 and 2, individual molecule of Imidazol, the optimal kernel found with the BIC method for 5 000 energy points was,
\begin{eqnarray}
 k_{1}(\mathbf{x}_i,\mathbf{x}_j) =  k_{2}(\mathbf{x}_i,\mathbf{x}_j) =a_0 \kMat + a_1\kMatt + a_2\kRbf .\nonumber \\
  \label{eqn:k_Hiroki_1}
\end{eqnarray}
The third fragment describes the bridge between fragments 1 and 2. The optimal kernel optimized by the BIC method is,
\begin{eqnarray}
 k_{12}(\mathbf{x}_i,\mathbf{x}_j) =  \left (a_0 \kMat \times \kRQ  + a_1\kRbf \right )\times \kMatt. \nonumber \\
  \label{eqn:k_Hiroki_12}
\end{eqnarray}

We found that, for $\dSD = $ 15 000 and $N = $ 15 000 we achieved a total RMSE of 0.056 kcal/mol, where for fragments 1 and 2 the average error is 0.032 kcal/mol and an error of 0.086 kcal/mol for the fragment that describe the interaction between both Imidazole molecules.
For 5 000 training points, the $\kSD$ kernel can interpolate a more accurate PES than with a GP with the Matern kernel, even with 3 times more training points; Table \ref{tbl:Imidazole}.

\begin{table}
  \caption{The RMSE of each fragment and total PES with 10,000 test energy  points spanned in the same energy range from the training data, as a function of the spectral, $\dSD$, and training points $N$. }
  \label{tbl:Imidazole}
  \begin{tabular}{ c | c | c c c | c}
     &   & \multicolumn{4}{c}{RMSE [kcal/mol]}  \\ 
    $\dSD$  &  $N$  & $\boldsymbol{\xi}_{1}$ & $\boldsymbol{\xi}_{2}$ & $\boldsymbol{\xi}_{12}$ & Total \\
    \hline
 \multirow{4}{3em}{10 000} & 5000  &   0.074 & 0.064 & 0.205 & 0.131 \\
 &10000  &   0.048 & 0.041 & 0.117 & 0.077 \\
 &15000  &    0.041 & 0.038 & 0.087 & 0.059 \\
 & 20000  &    0.062 & 0.039 & 0.073 & 0.060 \\     \hline
  \multirow{3}{3em}{15 000} & 5000  &   0.073 & 0.064 & 0.206 & 0.131 \\
 & 10000  &  0.044 & 0.038 & 0.113 & 0.073 \\
 & 15000  &  0.034 & 0.030 & 0.086 & \bf{0.056} \\ \hline  \hline
 $\kMat$ & 15000  & 0.227 & 0.162 & 0.102 & 0.171 \\ 
  $k_{1}, k_{2}, k_{12}$\textsuperscript{\emph{a}} & 5000  & 0.124 & 0.099 & 0.124 & 0.116 \textsuperscript{\emph{b}}\\  \hline
  \end{tabular}
  
  \textsuperscript{\emph{a}} Eqs. (\ref{eqn:k_Hiroki_1} -- \ref{eqn:k_Hiroki_12}).\\
  \textsuperscript{\emph{b}} From Ref. \cite{Hiroki_JCP}, the RMSE is 0.1815 kcal/mol.
\end{table}

\begin{figure}[h!]
\centering
\includegraphics[width=0.35\textwidth]{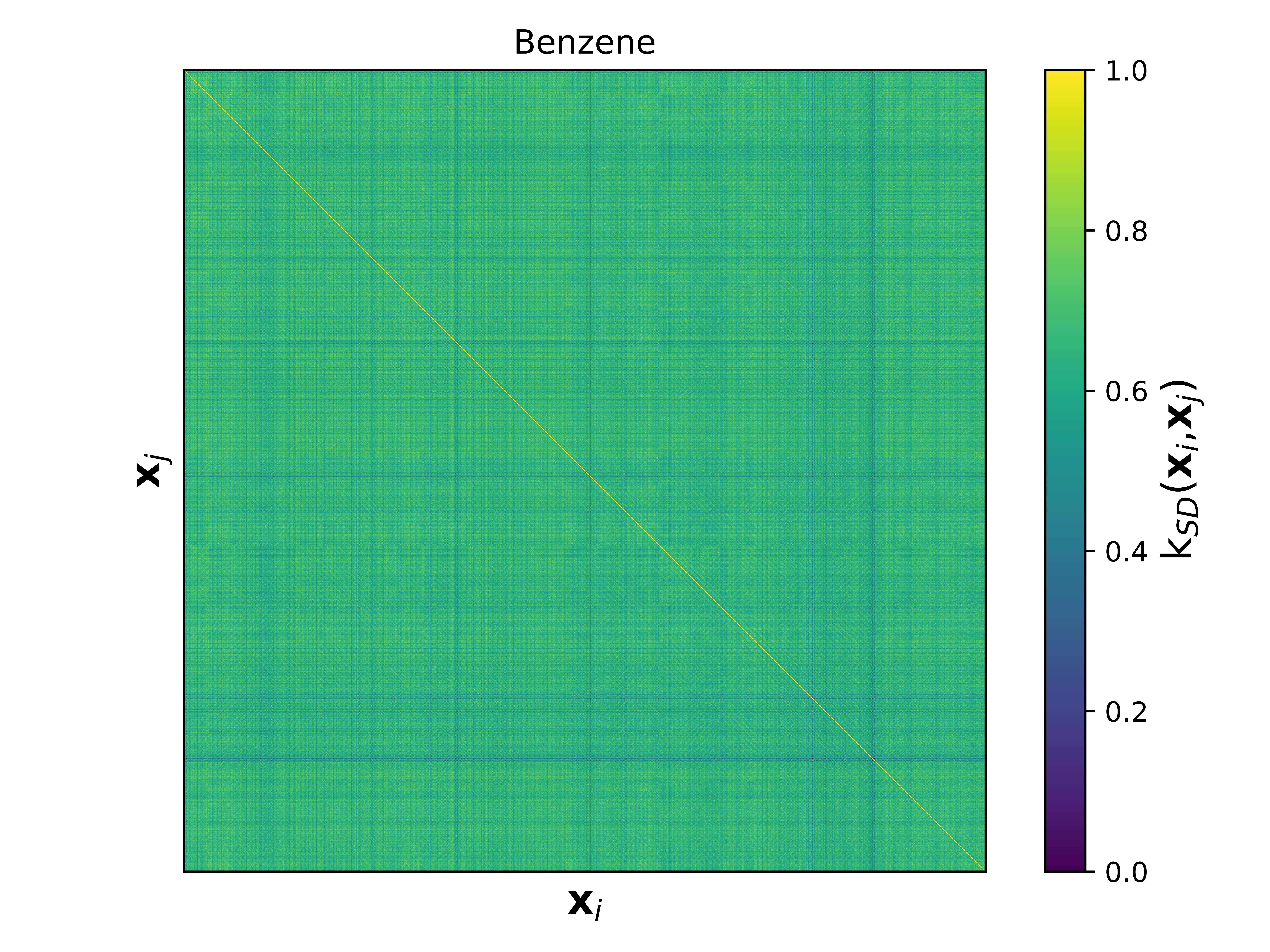}
\includegraphics[width=0.35\textwidth]{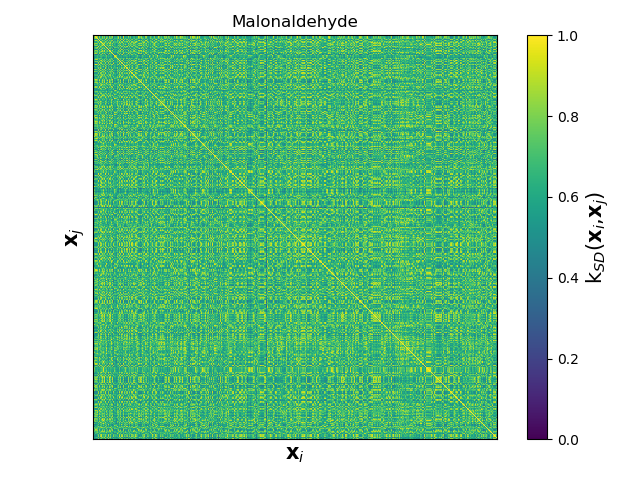}\\
\includegraphics[width=0.35\textwidth]{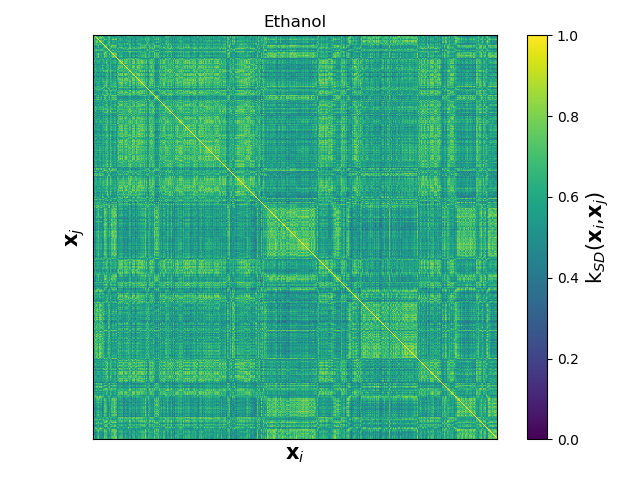}
\caption{For each molecule, we computed the value of spectral density covariance function, $\kSD$, for each pair of geometries ($\mathbf{x}_i,\mathbf{x}_j$) in the data set\cite{GDML_1,GDML_2}. $\kSD$ was optimized by maximizing the LMLL, Eq. \ref{eqn:lmll}, and the number of deltas for each system is reported in Table \ref{tbl:gpsd_qml7}. The displayed values were normalized for convenience.}
\label{fig:qml7_cov}
\end{figure}

We alaso consider the interpolation of the Benzene, Malonaldehyde, Ethanol, and Aspirin systems \cite{GDML_1,GDML_2}.
Our results illustrate that GPs, trained with the spectral density kernel, interpolate with high accuracy the PESs for high-dimensional molecular systems.  
For all four systems, we computed the mean absolute error (MAE) (Eq. \ref{eqn:MAE}) for the entire data set of points, including the training set. 
\begin{eqnarray}
\text{MAE} = \frac{1}{n} \sum^{n}_i  | y_i - \hat{y_i}|
\label{eqn:MAE}
\end{eqnarray}
All calculations for GPs were carried in a single GPU, Tesla T4 16GB. 
In Table \ref{tbl:gpsd_qml7}, we report the number of training and spectral points that lead to the most accurate GP. For all four systems, except for Malonaldehyde, we found that GPs with $\kSD$ interpolates more accurately that state-of-the-art deep-learning models; Table \ref{tbl:qml7}. We compared GPs with the $\kSD$ with deep-learning methods, e.g., deep-tensor NN (DTNN) \cite{DTNN}, PhysNet \cite{PhysNet}, and Cormorant \cite{Cormorant}, and KRR methods combined with gradients, e.g., GDML \cite{GDML_1,GDML_2}, and sGDML \cite{sGDML}.
Figure \ref{fig:qml7_cov} displays the value of the optimized $\kSD$ for each pair of geometries in the data sets for all four molecules. 

Aspirin is the largest system considered here, a 57D PES described with 210 features.
We found that with 15 000 training points and 2 000 spectral points, the MAE of this GP is  0.127 kcal/mol. By increasing both, $N$ and $|\delta|$ 	we managed to reduce the error to 0.063 kcal/mol, Table \ref{tbl:gpsd_qml7}.

\begin{table}
  \caption{The lowest error for four different molecular systems computed with a GP using the $\kSD$ kernel. The total error was computed for the entire data set\cite{GDML_1,GDML_2}, including the training points. }
  \label{tbl:gpsd_qml7}
  \begin{tabular}{ c | c  c | c c | c c  }
     &  \multicolumn{2}{c |}{$\kSD$}  & \multicolumn{2}{c |}{MAE}  & \multicolumn{2}{c}{RMSE}  \\ 
     & $N$ & $\dSD$ & [meV]  & [kcal/mol] & [meV]  & [kcal/mol]  \\  \hline
     Benzene & 10 000 & 5 000 & 0.31 & 0.0071 & 0.41 & 0.0094 \\ 
     Malonaldehyde & 20 000 &  5 000 & 2.91 & 0.067 & 4.44 & 0.102 \\
     Ethanol & 20 000 & 5 000 & 1.87 &  0.043 & 4.16 & 0.096\\  
     Aspirin & 20 000 & 5 000 & 2.73 & 0.063 & 3.65 &  0.0841 \\\hline  \hline
  \end{tabular}
  
\end{table}

\begin{table}
  \caption{MAE, reported in kcal/mol, for different ML models. The total error was computed for the entire data set\cite{GDML_1,GDML_2}, including the training points.   }
  \label{tbl:qml7}
  \begin{tabular}{ c | c  c | c }
 		&     $N$ [$10^{3}$]     &      ML model           &    MAE [kcal/mol] \\\hline
 \multirow{3}{3em}{Benzene} &   50   &   SchNet\textsuperscript{\emph{a}}       &  0.070 \\
 &   50   &   Cormorant\textsuperscript{\emph{b}}  &  0.020 \\
 &   10   &   GP $\kSD$($\dSD = $ 5K)   &  \bf{0.007} \\ \hline
  \multirow{4}{7em}{Malonaldehyde} &   15   &   GDML\textsuperscript{\emph{c}}       &  0.08 \\ 
 &   15   &  sGDML\textsuperscript{\emph{d}}      &  0.074 \\ 
 &   15   & PhysNet\textsuperscript{\emph{e}}     &  \bf{0.072} \\ 
 &   15   & GP $\kSD$($\dSD = $ 6K) &  0.079 \\ \hline
  \multirow{4}{7em}{Ethanol} &   15   &   GDML\textsuperscript{\emph{c}}       &  0.058 \\ 
 &   15   &  sGDML\textsuperscript{\emph{d}}      &  0.051 \\ 
 &   15   &  PhysNet\textsuperscript{\emph{e}}    &  0.050 \\ 
 &   15   & GP $\kSD$($\dSD = $ 6K) &  \bf{0.049} \\ \hline
 \multirow{4}{7em}{Aspirin} & 15 & \multirow{2}{4em}{GDML\textsuperscript{\emph{c}}} & 0.151 \\
 &   50   &     & 0.130 \\
 &   15   &  sGDML\textsuperscript{\emph{d}}  & 0.131 \\
 &   15   &  \multirow{2}{4em}{PhysNet\textsuperscript{\emph{e}}} & 0.124 \\
 &   50   &   & 0.121 \\ 
 &   50   &  SchNet\textsuperscript{\emph{a}} & 0.12 \\
 &   50   & Cormorant\textsuperscript{\emph{b}} & 0.098\\
 &   20   & GP $\kSD$($\dSD = $ 5K) &  \bf{0.063} \\ \hline  \hline
   \end{tabular}
  
  \textsuperscript{\emph{a}} Ref. \cite{SchNet}, 
  \textsuperscript{\emph{b}} Ref.\cite{Cormorant}, 
  \textsuperscript{\emph{c}} Refs.\cite{GDML_1,GDML_2},  
  \textsuperscript{\emph{d}} Ref.\cite{sGDML},
  \textsuperscript{\emph{e}} Ref.\cite{PhysNet}, 
\end{table}

\begin{figure}[h!]
\centering
\includegraphics[width=0.5\textwidth]{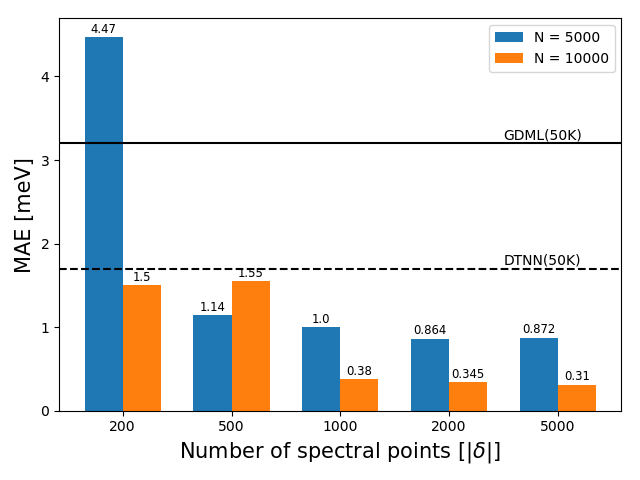}
\caption{MAE of the Benzene surface computed with a GP with a $\kSD$ kernel. We considered different number of spectral points.
We also considered different number of training points, $N=[5000,10000]$, displayed with different colors.
The horizontal solid line indicates the MAE of GDML \cite{GDML_1,GDML_2}, MAE = 3.2 meV, and the dashed line is the MAE of DTNN, MAE = 1.7 meV.
}
\label{fig:Benzene}
\end{figure}

For Benzene, we achieved a RMSE = $0.31$ meV with a GP trained with only 10 000 points and 5 000 spectral points.
However, a GP with only $N$ = 5 000 and  $|\delta| = 200$ is capable of predicting a more accurate PES than GDML and DTNN.
All results are displayed in Fig. \ref{fig:Benzene}.
For only 800 training points, a GP's MAE is 8.13 meV, with 500 spectral points; while the MAE of GDML with 1 000 points is 3.0 meV.
We found that, in the limit of low number of training points, a GP's accuracy is not comparable with models like GDML or sGDML. 

\begin{figure}[h!]
\centering
\includegraphics[width=0.5\textwidth]{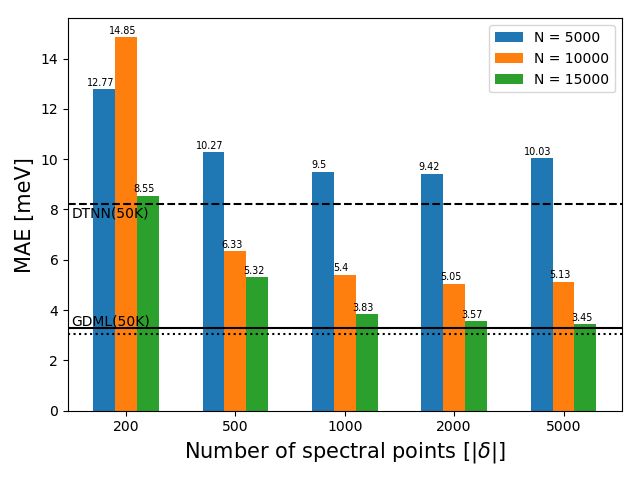}
\caption{MAE of the Malonaldehyde surface computed with a GP with a $\kSD$ kernel. We considered different number of spectral points.
We also considered different number of training points, displayed with different colors.
The dotted horizontal line is the MAE computed with a GP with $N = $ 18 000 and $|\delta| = $ 2 000, MAE = 3.05 meV.
The horizontal solid line indicates the MAE of GDML \cite{GDML_1,GDML_2}, MAE = 3.3 meV, and the dashed line is with DTNN, MAE = 8.2 meV.
}
\label{fig:Malonaldehyde}
\end{figure}

The results for Malonaldehyde are presented in Fig. \ref{fig:Malonaldehyde}.
For a GP trained with 15 000 points and 5 000 spectral points, the PES's accuracy is almost the same as the one predicted with GDML  \cite{GDML_1,GDML_2}.  However, by increasing the number of training points to 18 000, we achieved a more accurate prediction, MAE = 3.05, 0.35 meV more accurate than the one with GDML. 
Any PES interpolated with a GP trained with $\dSD \ge $ 500  and $N = $ 10 000 is more accurate than the one with DTNN.

\begin{figure}[h!]
\centering
\includegraphics[width=0.5\textwidth]{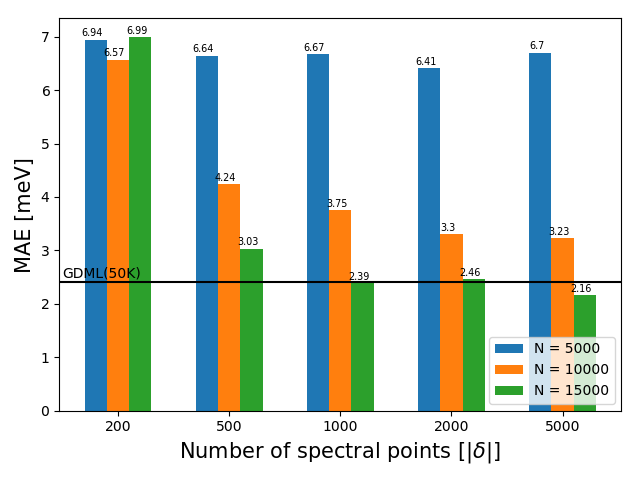}
\caption{MAE of the Ethanol surface computed with a GP with the $\kSD$ kernel considering different number of spectral points.
We also considered different number of training points,  displayed in different colors.
The horizontal solid line indicates the MAE of GDML \cite{GDML_1,GDML_2}, MAE = 2.4 meV.
}
\label{fig:Ethanol}
\end{figure}

For Ethanol, the predicted surface of a GP with $N = $ 15 000 and 1 000 spectral points is comparable with the one computed with GDML, see Fig. \ref{fig:Ethanol}.
However, a GP with a larger number of $\dSD$ is capable of interpolating this system more accurately; for a GP with $\dSD = $ 6 000 and $N = $ 15 000 points, the MAE is still the same as the one with 5 000 spectral points; MAE = 2.16 meV.

\section{Outline}
We have presented an accurate GP model capable of interpolating high dimensional PESs, i.g., 51D for protonated imidazole dimer, 61D for Benzene,  9D for both Malonaldehyde and Ethanol (36 features for both), and 57D for Aspirin surfaces (210 features). 
While combining simple kernels have proven to be a successful route, here we show an alternative path to enhance the accuracy of GPs based on the Bochners’ theorem.
This methodology can lead to highly accurate GPs capable of interpolating PESs for a variety of chemical systems.
The spectral delta kernel (Eq. \ref{eqn:k_SD}),  derived from the Bochners’ theorem, can also be optimized by maximizing the log-marginal likelihood as is commonly done for vanilla GPs.

For the protonated Imidazole dimer, we managed to predict a PES with a RMSE of 0.22 kcal/mol using 15 000 training and 2 000 spectral points. 
By splitting the total PES into three fragments, we achieved a PES with a total RMSE of 0.06 kcal/mol with $\dSD = [5 000, 10 000]$.
We found that $\kSD$ is only more accurate than the BIC method or simple kernels for systems with more than 5 000 points. 
Additionally, the optimal value of $\dSD$ depends on the number of training points too; for example, for Benzene $\dSD \ge $ 1 000 with N = 10 000 produce surfaces with MAEs lower than 1 meV ($\approx 0.023$ kcal/mol). 
The largest dimensional system considered here was Aspirin. For this system, we managed to predict a global surface with a MAE lower than 0.07 kcal/mol with 20 000 points, which compared with deep-learning models is more accurate. 
By eye-inspection, we found that the optimal $\dSD$ is between $\dSD = [N/3,N/2]$. 
A possible future work is to study how well GPs with the $\kSD$ kernel can extrapolating quantum observables, e.g., the high energy points for PES.

In physical sciences, the interpolation of high-dimensional landscapes is not commonly done by GPs unless the full covariance matrix is approximated by a low-rank matrix, for example using the Nystr\"om approximation, or using deep-NN which architecture must be optimize for different systems. 
Here, we illustrate that by combining modern deep-learning libraries such as GPytorch and KEOPS with GPUs, GPs are robust supervised ML algorithms capable of approximating high-dimensional complex functions without approximating the covariance matrix and still being training points efficient.
The work presented here makes GPs more suitable ML algorithms to study and simulate a wider variety of physical systems.

\begin{acknowledgements}
We thank R. Krems and H. Sugisawa for useful discussions.
This work was partially supported by the U.S. Air Force
Office of Scientific Research (AFOSR) in a grant,
FA9550-20-1-0354, to Professor P. Brumer, University
of Toronto.
\end{acknowledgements}

\bibliography{references}

\end{document}